\RequirePackage{ifpdf}
\documentclass[twoside]{JINST}
\pdfoutput=1
\usepackage{graphicx}
\usepackage{rangecite}
\usepackage{amsmath}
\usepackage[figuresright]{rotating}
\usepackage{afterpage}
\usepackage{float}
\usepackage{footnote}
\makesavenoteenv{tabular}

\title{Development of a camera casing suited for cryogenic and vacuum applications}

 \author{S.~C.~Delaquis\thanks{Corresponding author}, R.~Gornea, S.~Janos, M.~L\"uthi, Ch.~Rudolf~von~Rohr, M.~Schenk, and J.-L.~Vuilleumier \\
 \llap{}Albert Einstein Center for Fundamental Physics, \\
 Laboratory for High Energy Physics, \\
 University of Bern, \\
 Sidlerstrasse~5,\\
 CH-3012 Bern, Switzerland \\
 E-mail: \email{delaquis@LHEP.unibe.ch}
}

\abstract{
We report on the design, construction, and operation of a PID temperature controlled and vacuum tight camera casing. The camera casing contains a commercial digital camera and a lighting system. The design of the camera casing and its components are discussed in detail. Pictures taken by this cryo-camera while immersed in argon vapour and liquid nitrogen are presented. The cryo-camera can provide a live view inside cryogenic set-ups and allows to record video.
}


\begin{document}

\section{Introduction}

Digital imaging is a field where fast progress has been made. In experimental physics digital cameras often need to function under especially rough and demanding conditions, such as, extreme temperatures, wide pressure range, high magnetic fields, strong ionizing radiation, etc. A digital camera working in a cryogenic and high vacuum environment is needed for the R\&D program on barium ion tagging performed by the EXO (Enriched Xenon Observatory) group in Bern~\cite{Danilov}. The EXO-100 cryostat is a R\&D set-up to test the feasibility of extracting single ions from a Time Projection Chamber (TPC) filled with liquid xenon. For the extraction of ions a mechanical device is under development. To monitor this device a camera is installed inside the cryostat. This camera has to withstand low temperatures and allow for high vacuum. However, few camera models could be found to meet these demands. Therefore, a custom temperature controlled camera casing was developed to house a conventional digital camera.

\section{Materials and methods}

\subsection{Digital camera}

The digital camera used here is a C525 model manufactured by Logitech. This model was chosen because of its compact design and low price. To further minimize its size the stand was removed. Its most important specifications are summarized in Table~\ref{tab:CamSpecs}.

\begin{table}[ht]
\renewcommand{\arraystretch}{1.25}
\begin{center}
\begin{tabular}{|l|l|l|}
  \hline
  HD video capture & up to 1280 x 720 pixels\\
  Photo capture & up to 8 megapixels (software enhanced)\\
  Audio & built-in mic\\
  Interface & Hi-Speed USB 2.0\\
  \hline
 \end{tabular}
\caption[]{Specifications of the Logitech HD Webcam C525 \footnotemark[1].}
\label{tab:CamSpecs}
\end{center}
\end{table}

\subsection{PID temperature regulation}

\footnotetext[1]{Logitech; www.logitech.com/en-ch/product/hd-webcam-c525}
\setcounter{footnote}{1}

\begin{figure}[H]	
	\center\includegraphics[width=0.9\textwidth]{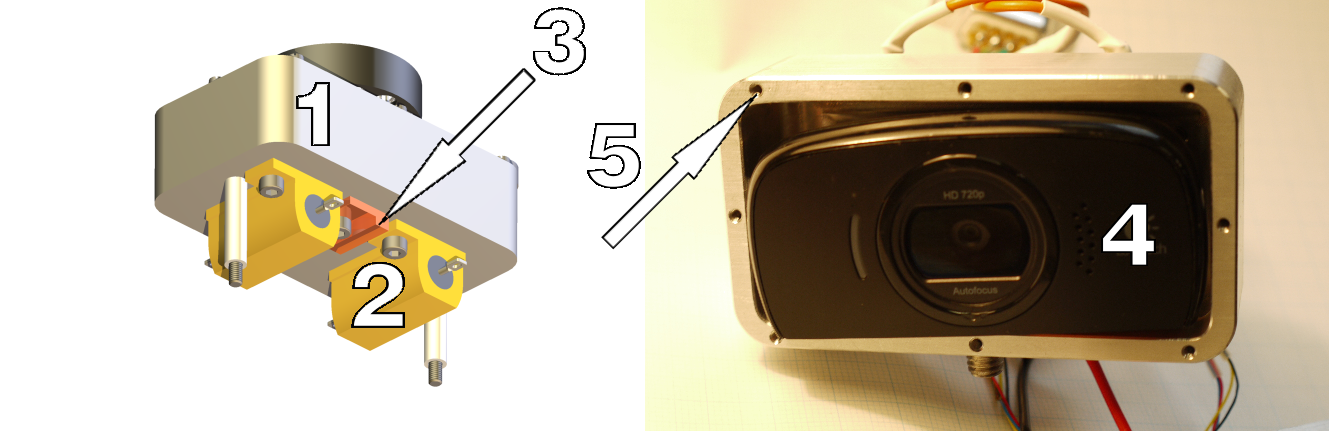}
	\caption{Left: CAD render, worm's eye view of the inner unit: 1- inner aluminium casing, 2- two heating resistors, 3- copper support for PT-100 sensor.
	Right: photograph of the inner casing with removed lid: 4- camera with removed stand, 5- threads to close the lid.}
	\label{fig:rd_ph_inner}
\end{figure}

To keep the camera within its operating temperature range (between -30$^\circ$C and +50$^\circ$C), active temperature regulation is needed. The ITC-503 temperature controller unit from Oxford Instruments\footnote{Oxford Instruments; http://www.oxford-instruments.com} is used together with a platinum-100 sensor (PT-100) and two heating resistors (each 39~Ohm and 25~W, in parallel) to form a PID-controlled temperature regulation system. Good thermal contact between camera, PT-100, and heating resistors is required for the temperature regulation system to work. Therefore, the camera is inserted in a small aluminium casing with a lid (fig.~\ref{fig:rd_ph_inner}). The choice fell on aluminium because of its good thermal conductivity. The PT-100 - inserted into a copper support - and the heating resistors are attached with screws to the bottom of the aluminium casing, guaranteeing a good thermal contact. This inner unit is maintained at a constant temperature by the ITC-503. The target temperature and the PID-parameters are set at the ITC-503 to $T=250$~K, $P=10.0$, $I=1.5$, and $D=0.1$.

\begin{figure}[H]	
	\center\includegraphics[width=0.35\textwidth]{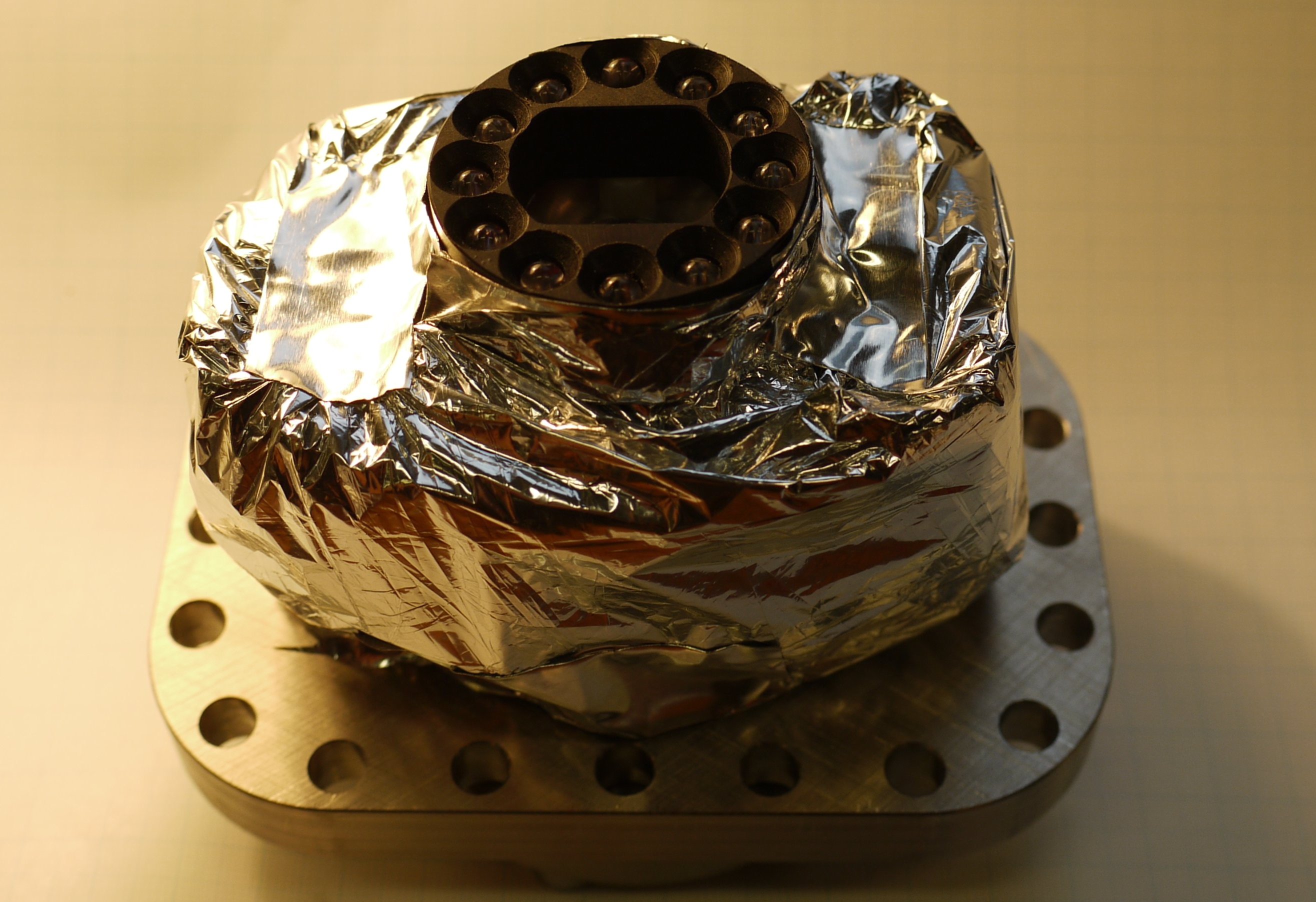}
	\caption {Photograph of the inner unit wrapped in five layers of super insulation.}
	\label{fig:ph_cam-supr-insul}
\end{figure}

To regulate the temperature of the inner unit the ITC-503 provides power to the two heating resistors. Ultimately, the energy is dissipated to the environment and should therefore be kept small. The required heating power can be minimized by thermally insulating the inner unit from the environment. Therefore, special attention was paid to minimizing the heat radiation. The inner aluminium casing forms a heat radiation barrier between the environment and the camera. The barrier is only broken for the lens of the camera. To reduce the heat exchange between the inner unit and the outer body, the inner unit is completely wrapped in five layers of super insulation (metallised boPET film). A small hole is left at the position of the lens (fig.~\ref{fig:ph_cam-supr-insul}).

\subsection{Vacuum tight camera body}

\begin{figure}[H]	
	\center\includegraphics[width=0.9\textwidth]{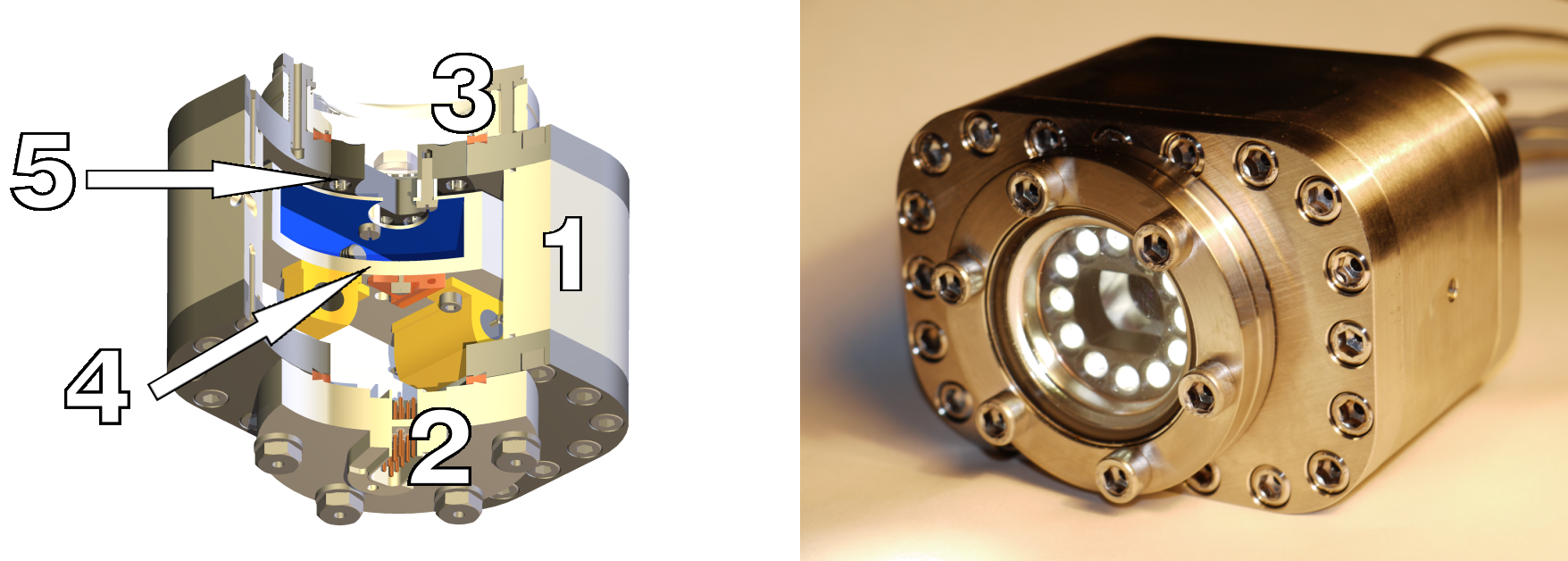}
	\caption{Left: worm's eye section view of the cryo-camera: 1- vacuum tight camera body, 2- Sub-D 9 pin feed-through, 3- quartz window, 4- inner unit with space for the camera (blue), 5- lighting system. Right: photograph of the cryo-camera.}
	\label{fig:rd_cut-bottom}
\end{figure}

The main application of the cryo-camera is its operation inside the EXO-100 cryostat. The EXO-100 cryostat can be filled with diverse gases and liquids. Electronegative impurities dissolved in those liquids must be reduced down to a level of less than 1~ppb O2 equivalent. To reach such purity levels the cryostat is evacuated to high vacuum prior to filling it with a pure medium. However, the camera's out gassing rate is expected to be severe due to its plastic frame and printed circuit board (PCB). A high out gassing rate makes it almost impossible to reach such low pressures and certainly contaminates the pure medium. Therefore, the inner unit - containing the camera - has to be encapsulated in a camera body with a low out gassing rate (fig.~\ref{fig:rd_cut-bottom}).

\begin{figure}[H]	
	\center\includegraphics[width=0.9\textwidth]{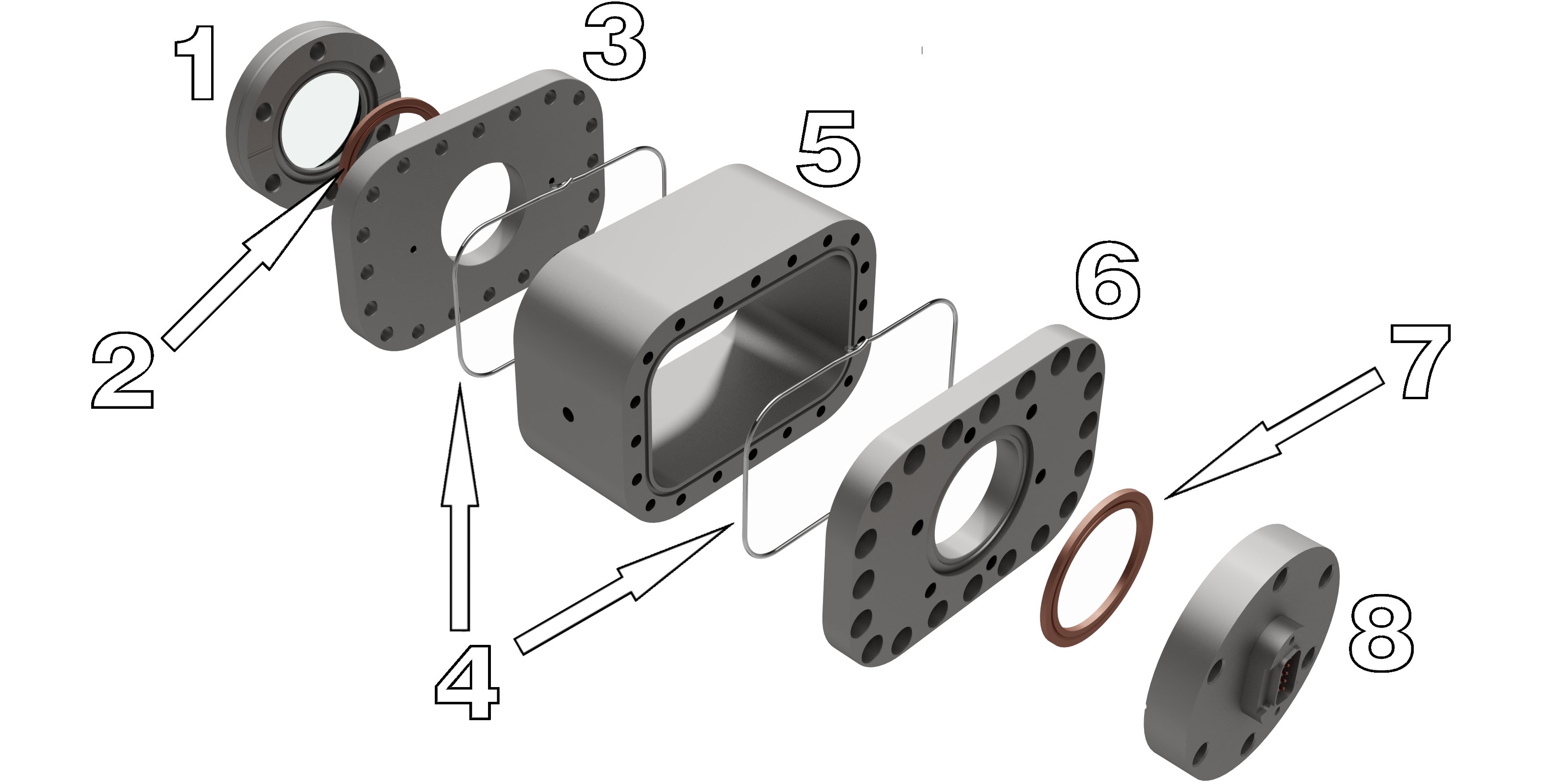}
	\caption{CAD exploded view of the vacuum tight camera body (screws are omitted): 1- quartz window on CF-40 , 2- soft baked copper gasket, 3- upper flange, 4- indium wires, 5- body piece, 6- lower flange, 7- copper gasket, 8- Sub-D 9 pin feed-through on CF-40.}
	\label{fig:rd_explosion}
\end{figure}

The camera body is vacuum tight and built out of materials well suited for high vacuum (e.g. low out gassing and gas permeation). To thermally insulate the inner unit from the cryogenic environment materials with low heat conductivity are preferred candidates (e.g. polymers such as PAI or PEEK, or metals such as stainless steel or titanium). A compromise between high vacuum capability and thermal insulation had to be found. In this case, stainless steel was chosen due to its good high vacuum capability, moderate heat conductivity, low price, and good machinability. The camera body's parts are shown in Figure~\ref{fig:rd_explosion}.

\begin{figure}[H]	
	\center\includegraphics[width=0.4\textwidth]{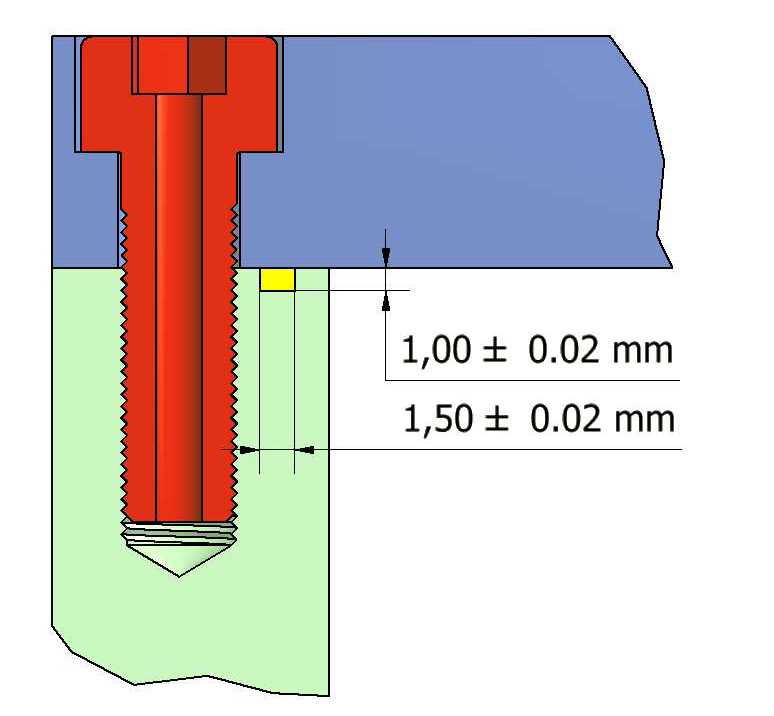}
	\caption{CAD detail of the indium sealing: blue- flange, turquoise- body piece, yellow- groove for indium wire, red- vacuum screw.}
	\label{fig:rd-vacuum}
\end{figure}

To connect the individual parts together two well known vacuum sealing techniques are used~\cite{ARoth}. A quartz window and a Sub-D 9 pin feed-through are mounted to the flanges with a CF-40 soft baked copper gasket and a normal copper gasket, respectively. Because of the CF's cutting edge the flanges have to be made out of a hard metal. The seals between the flanges and the body piece are made with indium wires~\cite{Richardson}. A 1.5~mm wide and 1.0~mm deep groove is milled on both sides of the body piece (fig.~\ref{fig:rd-vacuum}) with a slightly smaller cross section than that of a 1.5~mm diameter indium wire. This design guarantees that the indium flows along both sealing surfaces while they are being pressed together and thus forms a tight seal. Furthermore, a thin film of APIEZON N vacuum grease is put on the sealing surfaces before the indium wires are put into the grooves. The grease fills small scratches on the sealing surfaces and will allow for easy removal of the indium once the seal has to be opened. To avoid virtual leaks vacuum screws are used.

\subsection{Lighting system}

When mounted inside the EXO-100 cryostat a lighting system is needed to illuminate the scene. Twelve LED's are mounted on a support ring made of POM (polyoxymethylen). This ring is fixed onto the inner unit inside of the vacuum tight camera body (fig.~\ref{fig:rd_cut-bottom}). In the middle of the support ring a hole was cut out to leave the view of the camera free. The LED's can be dimmed or turned off from an external control unit.

\subsection{Gas purification system}

The gas filling the inner volume of the camera body may condense or freeze on the camera body's inner walls once it is cooled down to cryogenic temperature. Therefore, there are the risks that the pressure in the inner volume drops significantly (which could lead to overheating of the camera) and that the quartz window becomes misted. To avoid these problems, an adequate gas is chosen to fill the inner volume and a simple gas purification system is included.

The choice for a filling gas depends on the cryo-camera's application (expected coldest body temperature) and is lead by a compromise between good thermal insulation (e.g. argon, xenon, or nitrogen) and low condensation temperature (e.g. helium 4~K, nitrogen 77~K, argon 86~K, or xenon 165~K; at 1~bar). Even though helium is a very inefficient insulator, its low condensation temperature and the ability to easily leak check it, makes it the preferred candidate. Regarding thermal insulation nitrogen is the better choice for most cryo-camera's applications.

\begin{figure}[H]	
	\center\includegraphics[width=0.4\textwidth]{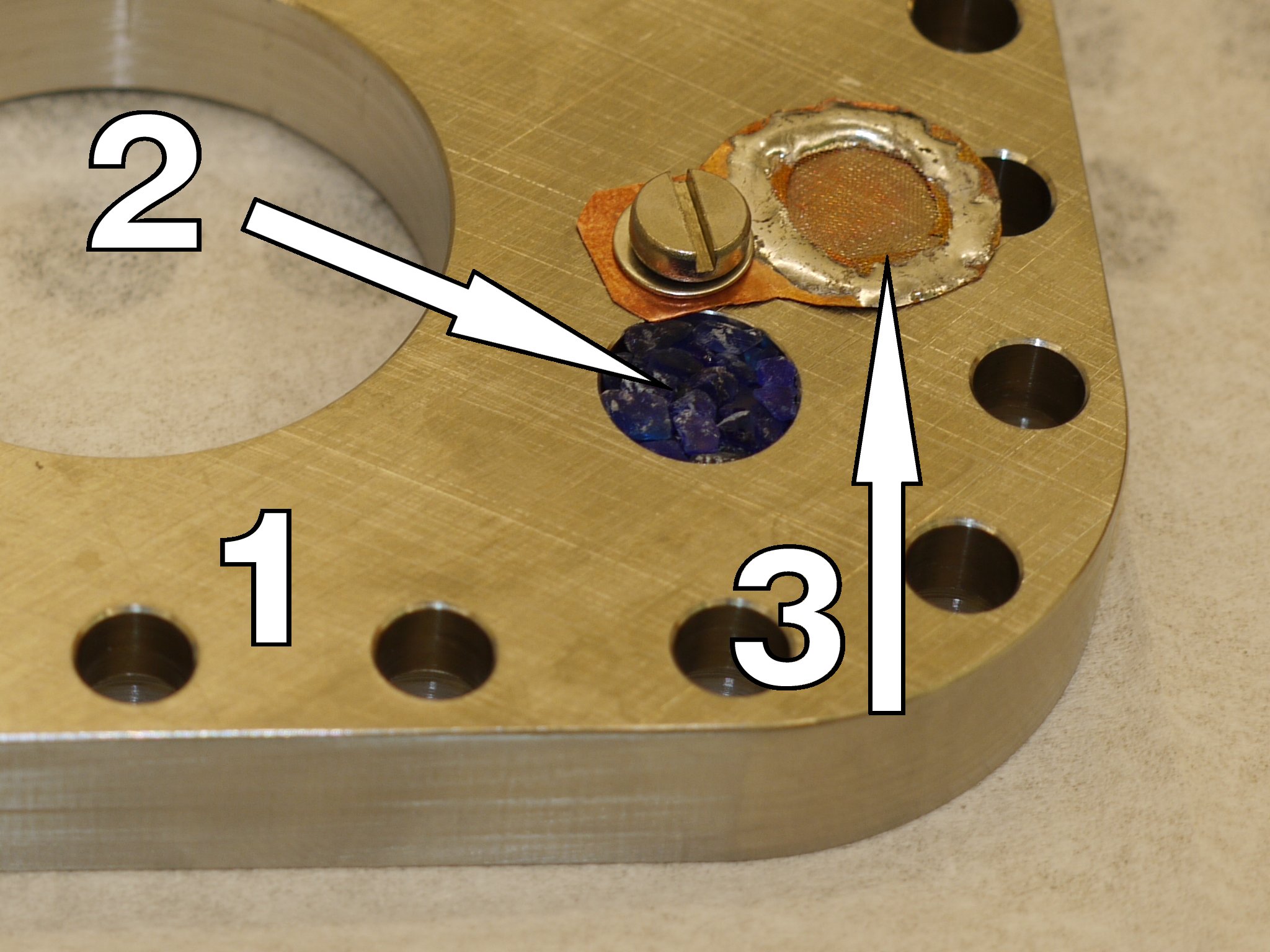}
	\caption{Photograph of the gas purifying system: 1- upper flange, 2- hole filled with silica gel, 3- cover with a thin mesh to close the hole.}
	\label{fig:ph_purity}
\end{figure}

Unwanted residual components with high freezing temperatures (e.g. water) are still present after the inner volume is filled with helium gas and can still mist the quartz window. These residuals originate from impurities in the filling gas and out gassing from the camera parts. A simple purification system is integrated in the camera body to remove these residuals from the filling gas. The purification system consists of two parts (fig.~\ref{fig:ph_purity}). Firstly, a hole is milled in the upper flange reducing locally the wall thickness to 0.5~mm and thus lowering locally the heat resistance. Therefore, this will be the coldest point inside the camera body once the cryo-camera is at cryogenic temperature. The residuals will condense at this point. Secondly, this hole is filled with silica gel to efficiently remove water vapour from the filling gas.

\newpage

\subsection{Sealing technique}

During the closing procedure, the Sub-D feed-through was removed from the camera body and a turbo molecular pump was installed instead. The inner volume was then evacuated for a minimum of 15 minutes to remove components with high vapour pressure. Thereafter, helium was used to break the vacuum. Quickly, the pump was removed and the feed-through was loosely reinstalled. In a next step, the cryo-camera was put into a small vacuum chamber, with the loosely attached feed-through facing downwards - so the helium could not escape during installation. The chamber was evacuated with a roughing pump and then flushed again with helium. This step was repeated twice. Finally, the cryo-camera was taken out and the feed-through was tightened, sealing the helium filled inner volume.

\section{Measurements and discussion}

\subsection{Vacuum tightness}

To test the vacuum tightness of the camera body the cryo-camera was placed inside a small vacuum chamber. The vacuum chamber was equipped with three devices: a turbo molecular pump to evacuate the chamber, a vacuum gauge to measure the pressure, and a helium leak checker (PHOENIXL 300 by \OE rlikon) to detect leaks in the camera body. A first test was done with the empty chamber. Thereafter, the chamber was reopened and the cryo-camera was installed. The results of the two runs are shown in Table~\ref{tab:CamVacuum}.

\begin{table}[ht]
\renewcommand{\arraystretch}{1.25}
\begin{center}
\begin{tabular}{ l | c | c }
   {\bf condition} & {\bf pressure [mbar]} & {\bf leak rate [mbar $\cdot$ l/s]}\\
  \hline
    empty &  $1.1 \cdot 10^{-6}$ & $1.4 \cdot 10^{-10}$\\
  \hline
    cryo-camera &  $1.7 \cdot 10^{-6}$ & $3.3 \cdot 10^{-10}$\\
 \end{tabular}
\caption[]{Results of the vacuum tightness test.}
\label{tab:CamVacuum}
\end{center}
\end{table}

\begin{figure}[H]	
	\center\includegraphics[width=0.4\textwidth]{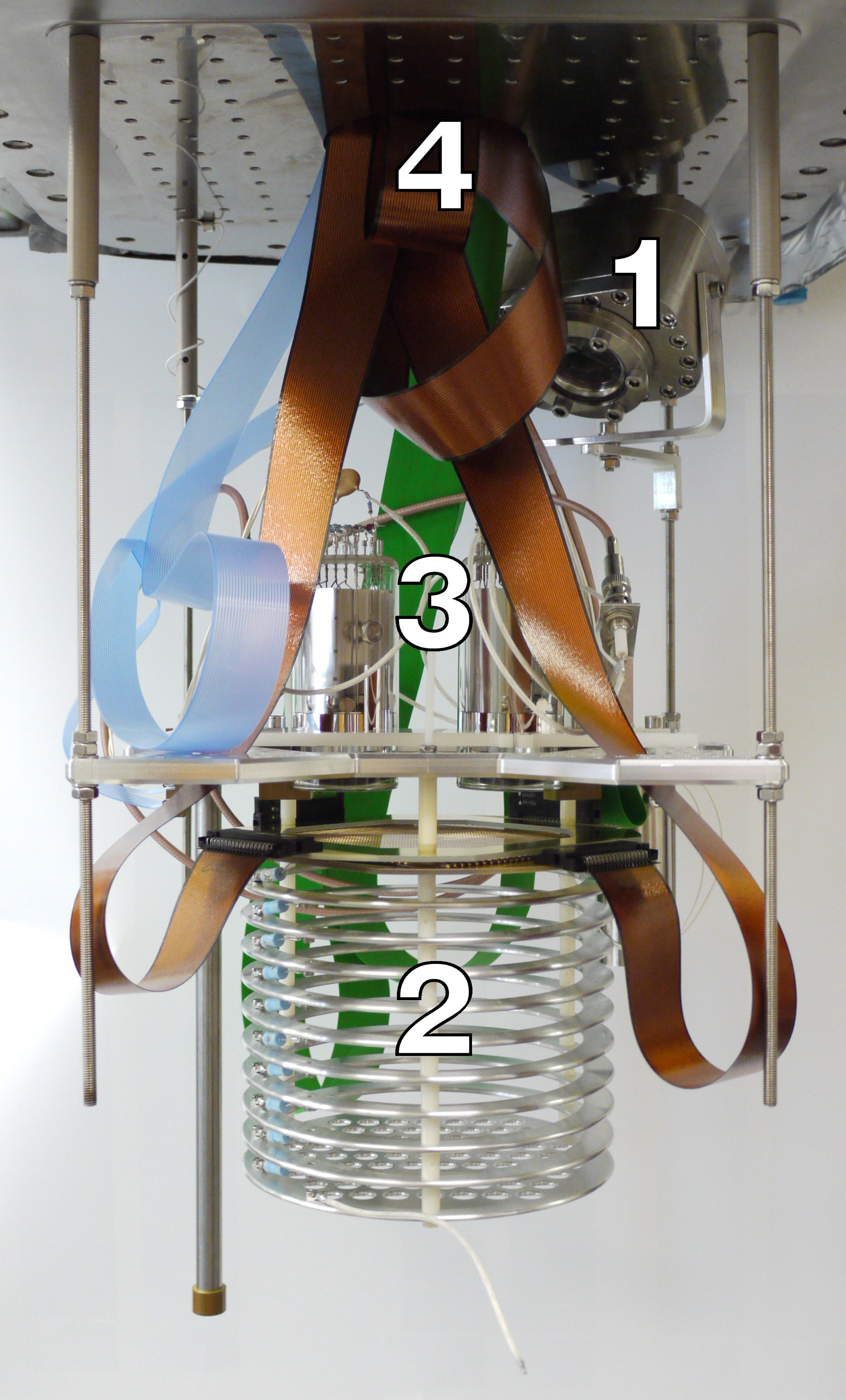}
	\caption{Photograph of the EXO-100 detector with the mounted cryo-camera: 1- cryo-camera, 2- time projection chamber, 3- photo multiplier tubes, 4- signal cables.}
	\label{fig:ph_cam-mounted}
\end{figure}

For operation in environments where high purity is required (e.g. TPC's) the values shown in Table~\ref{tab:CamVacuum} are sufficient. After these tests, the cryo-camera was operated inside the EXO-100 cryostat. The inner volume of the EXO-100 cryostat consists of two connected cubic chambers on top of each other. The lower chamber hosts a TPC with 15~cm drift gap, photo multiplier tubes (PMT's), and additional instrumentation. The upper chamber is reserved for a displacement device to extract single ions from the TPC. The cryo-camera is installed in the upper chamber facing towards the TPC, see Figure~\ref{fig:ph_cam-mounted}. During TPC operation the lower chamber is filled with liquid argon and the upper chamber is filed with argon vapour. The cryo-camera was successfully operated inside the cryostat while tracks of cosmic muons were observed with the TPC. The analysis of these tracks confirms that the cryo-camera does not pollute its environment.

\subsection{Temperature read-out and control}

\begin{figure}[H]	
	\center\includegraphics[width=0.9\textwidth]{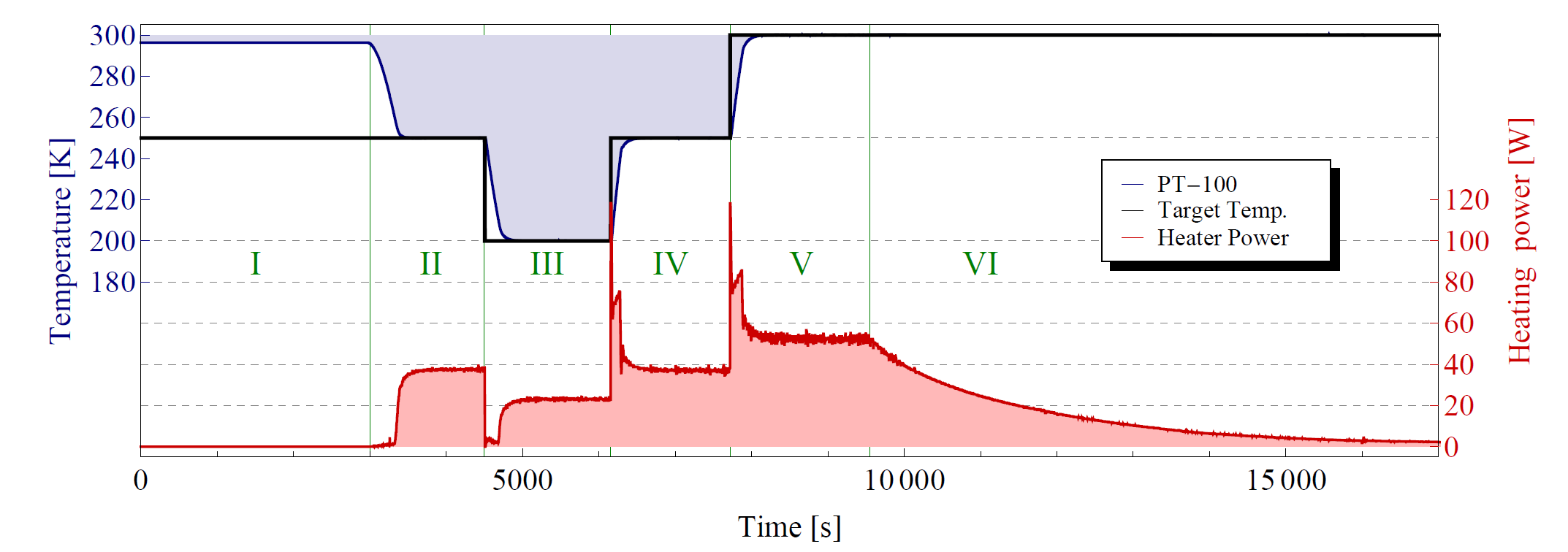}
	\caption{Temperature of the inner unit and heating power: I- cryo-camera at room temperature ($T_t=250$~K), II- cryo-camera immersed into liquid nitrogen ($T_t=250$~K), III- $T_t$ lowered to 200~K, IV- $T_t$ increased back to 250~K, V- $T_t$ increased to 300~K, VI- cryo-camera extracted from liquid nitrogen to air ($T_t=300$~K).}
	\label{fig:gr_t}
\end{figure}

Important parameters for applications of the cryo-camera are the operation temperature range, the temperature stability, and the dissipated power. To measure these quantities under demanding conditions the cryo-camera was immersed in liquid nitrogen (77~K). During this procedure, the ITC-503 measured the temperature of the inner unit (PT-100) and maintained it at a given target temperature $T_t$. A computer was connected to the ITC-503 (via RS-232 interface) to record the temperature curve of the inner unit and the voltage of the ITC-503's heating output. Thereafter, the heating power was calculated from the heating output voltage (over a total resistance of 19.5~Ohm). Figure~\ref{fig:gr_t} shows the recorded data.


In this test, target temperatures of the inner unit between 200~K and 300~K were reached. Below 250~K the contrast of the image was reduced and at 200~K the camera stopped working. However, the camera fully recovered after warming up the inner unit to 250~K. During earlier tests, the inner unit was cooled down to 77~K and the camera also fully recovered after warming up. Similar behaviour was reported by Seddon et al.~\cite{Seddon} who used a CCD camera to visualize cryogenic flow. At 250~K the camera works without any image degradation, and thus, this temperature was chosen as the nominal operation temperature.

The data from this test show a precise temperature regulation and an excellent temperature stability ($T = T_t \pm 0.1$~K). However, the absolute temperature accuracy depends on the sensor calibration at the ITC-503. In this case the uncertainty is $\pm 1$~K. Better accuracy can be reached by carefully calibrating the PT-100 sensor.

The required heating power $P_h$ to maintain the inner unit at a certain $T_t$ is shown in Table~\ref{tab:Power}. During camera operation ($T_t=250$~K) $38$~W are constantly dissipated into the liquid nitrogen. Hence, the nitrogen constantly boils and produces bubbles at the surface of the camera body. To reduce boiling, the target temperature can be set to a lower temperature while the camera is not in operation (e.g. $T_t=200$~K for fast recovering within minutes or $T_t<77$~K to avoid any boiling). This will significantly lower the power dissipated into the liquid nitrogen. Furthermore, operation of the cryo-camera in liquid nitrogen is considered a demanding condition, therefore, these values represent the worst case scenario. During operation in the vapour phase, as in the case of the EXO-100 detector, less heat is lost and no bubbles are produced in the liquid phase underneath the cryo-camera.

\begin{table}[ht]
\renewcommand{\arraystretch}{1.25}
\begin{center}
\begin{tabular}{ l | c | c | c }
   {\bf  $\mathbf{T_t}$ [K]} &  $300 \pm 1$ &  $250 \pm 1$ &  $200 \pm 1$\\
  \hline
   {\bf $\mathbf{P_h}$ [W]} &  $52 \pm 3$  &  $38 \pm 3$  &  $23 \pm 3$ \\
 \end{tabular}
\caption[]{Heating power $P_h$ required to maintain the inner unit at a certain target temperature $T_t$ (cryo-camera immersed in liquid nitrogen).}
\label{tab:Power}
\end{center}
\end{table}

\subsection{Digital imaging}

\begin{figure}[H]	
	\center\includegraphics[width=0.6\textwidth]{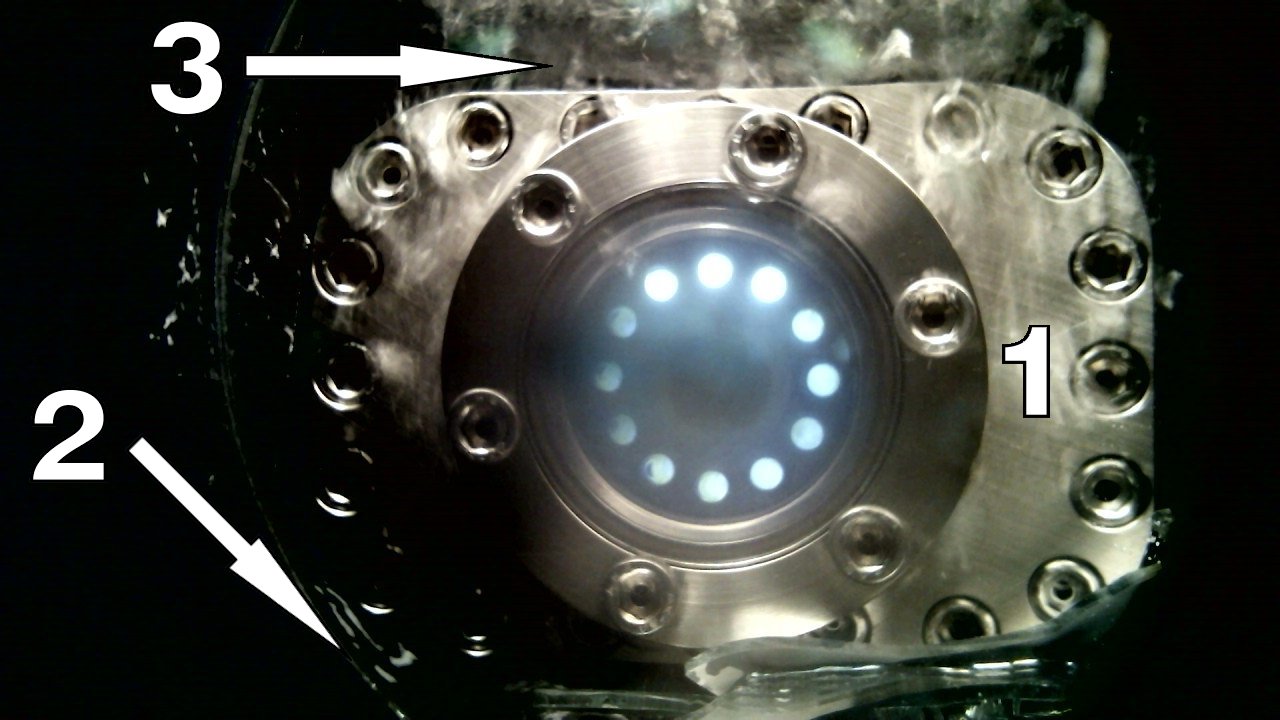}
	\caption{Picture taken with the cryo-camera while immersed in liquid nitrogen. A mirror was placed 10~cm in front of the cryo-camera and a self portrait was taken: 1- cryo-camera, 2- round three times magnifying shaving mirror, 3- bubbles produced by the cryo-camera due to heat dissipation}
	\label{fig:vw_portret}
\end{figure}

\begin{figure}[H]	
	\center\includegraphics[width=0.6\textwidth]{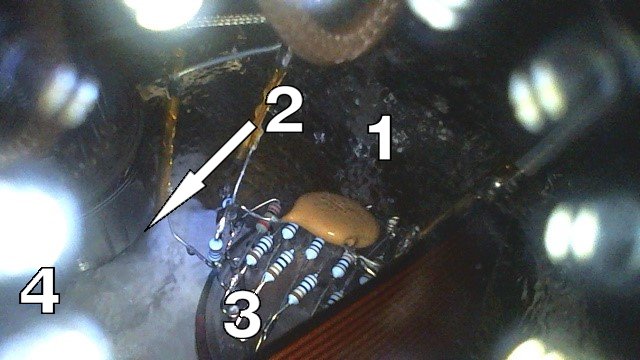}
	\caption{Picture taken with the cryo-camera while installed in the EXO-100 cryostat. Bird's eye view onto the EXO-100 TPC filled with liquid argon: 1- TPC immersed in liquid argon, 2- level of the liquid, 3- two photo multiplier tubes, 4- reflections of the front LED's on the quartz window.}
	\label{fig:vw_tpc2}
\end{figure}

The cryo-camera's imaging performance was tested under two conditions. In a first test the cryo-camera was exposed to short-term harsh conditions (for a few hours, immersed in liquid nitrogen). In a second test the cryo-camera was installed in the EXO-100 cryostat and operated under long-term mild conditions (for several weeks, above liquid argon in the vapour phase). In both cases the target temperature of the inner unit was set to $T_t=250$~K at the ITC-503. During these tests several pictures and videos were recorded with the cryo-camera. Two pictures are shown in Figures~\ref{fig:vw_portret}~and~\ref{fig:vw_tpc2}.

Under both conditions the cryo-camera performed well and no image degradation was observed. However, three problems occurred. Firstly, the light of the lighting system is reflected on the front window and disturbs the outer part of the camera's view field. A possible improvement would be to replace the current window with one having an anti-reflection coating. This would lower the intensity of the reflected light and thus reduce the undesirable effect. Secondly, sometimes impurities condensed and froze on the inside of the window during operation in liquid nitrogen. This reduces or blocks the view of the camera. Thirdly, the dissipated heat of the heating resistors induces boiling in liquid nitrogen. Therefore, the cryo-camera has to be oriented horizontally as otherwise the bubbles disturb the camera view field. The last two problems only occur during operation in liquid nitrogen (and most likely in liquid argon). Operation in the gas phase is free of these problems.


\section{Conclusion}

A camera casing was developed to allow for operation of inexpensive digital cameras in cryogenic and clean environments. The temperature of the camera is maintained within its operating range by a PID-loop. Furthermore, a lighting system is included in the camera casing to light up the camera's field of view. A prototype was operated successfully in liquid nitrogen and argon vapour. The limitations of this cryo-camera are clear: the rather large power consumption of the heating resistors leads to boiling of the surrounding cryogenic liquid. In addition, light of the lighting system is partially reflected on the front window, and thus, limits the camera's effective field of view. A front window with an anti-reflection coating would reduce this effect significantly. Notwithstanding these limitations, the cryo-camera is well suited for our application in the vapour phase. Moreover, one such cryo-camera was operated successfully in the LAPD (Liquid Argon Purity Demonstrator at Fermilab) to study high voltage discharges in liquid argon. The installation of several such cryo-cameras in the MicroBooNE detector has been considered.

The concept presented here could be extended to employ a camera sensitive to IR to allow for simultaneous operation of camera, lighting system (built with IR LED's), and photo multiplier tubes (sensitive only to UV and visible - wavelength of scintillation light in most noble liquids). Moreover, a high speed camera would allow to study the temporal evolution of electric discharges in cryogenic liquids.

\section*{Acknowledgements}
We wish to thank the LHEP technical staff for their very valuable help in machining the casing of the camera, namely R.~H\"anni, J.~Christen, F.~Nydegger, and R.~Liechti. This work was supported by the Swiss National Science Foundation.

\end{document}